\documentclass[12pt]{article}
\usepackage{amsmath}
\usepackage{psfrag,epsf,graphicx}
\usepackage{enumerate}
\usepackage{amsfonts}
\usepackage{amsmath}
\usepackage{listings}

\DeclareMathOperator*{\argmin}{arg\,min}
\usepackage{url} 
\usepackage[colorlinks,citecolor=blue,urlcolor=blue]{hyperref}
\usepackage[round]{natbib}  
\usepackage{algorithm,algorithmicx}
\usepackage{algpseudocode}
\algdef{SE}[SUBALG]{Indent}{EndIndent}{}{\algorithmicend\ }%
\algtext*{Indent}
\algtext*{EndIndent}

\newcommand{\blind}{0}

\begin{document}

\def\spacingset#1{\renewcommand{\baselinestretch}%
{#1}\small\normalsize} \spacingset{1}


\if0\blind
{
  \title{\bf Variable Selection with Second-Generation P-Values}
  \author{Yi Zuo\thanks{Yi Zuo is PhD candidate, Department of Biostatistics, Vanderbilt University, Nashville, TN 37203 (\href{mailto:yi.zuo@vanderbilt.edu}{yi.zuo@vanderbilt.edu}); Thomas G. Stewart is Assistant Professor, Department of Biostatistics, Vanderbilt University, Nashville, TN 37203 (\href{mailto:thomas.stewart@vanderbilt.edu}{thomas.stewart@vanderbilt.edu}); and Jeffrey D. Blume is Professor, Department of Biostatistics, Vanderbilt University, Nashville, TN 37203 (\href{mailto:j.blume@vanderbilt.edu}{j.blume@vanderbilt.edu})   }\hspace{.2cm}\\
    Department of Biostatistics, Vanderbilt University \\
    Thomas G. Stewart  \\
    Department of Biostatistics, Vanderbilt University \\
    and \\
    Jeffrey D. Blume \\
    Department of Biostatistics, Vanderbilt University
        }
  \maketitle
} \fi

\if1\blind
{
  \bigskip
  \bigskip
  \bigskip
  \begin{center}
    {\LARGE\bf Variable Selection with Second-Generation P-Values \par}
\end{center}
  \medskip
} \fi

\bigskip
\begin{abstract}
Many statistical methods have been proposed for variable selection in the past century, but few balance inference and prediction tasks well. Here we report on a novel variable selection approach called Penalized regression with Second-Generation P-Values (ProSGPV). It captures the true model at the best rate achieved by current standards, is easy to implement in practice, and often yields the smallest parameter estimation error. The idea is to use an $\ell_0$ penalization scheme with second-generation p-values (SGPV), instead of traditional ones, to determine which variables remain in a model. The approach yields tangible advantages for balancing support recovery, parameter estimation, and prediction tasks. The ProSGPV algorithm can maintain its good performance even when there is strong collinearity among features or when a high dimensional feature space with $p>n$ is considered. We present extensive simulations and a real-world application comparing the ProSGPV approach with smoothly clipped absolute deviation (SCAD), adaptive lasso (AL), and minimax concave penalty with penalized linear unbiased selection (MC+). While the last three algorithms are among the current standards for variable selection, ProSGPV has superior inference performance and comparable prediction performance in certain scenarios. Supplementary materials are available online.
\end{abstract}

\noindent%
\vfill

\newpage
\spacingset{1.45} 
\section{Introduction}
\label{sec:intro}
Data are typically comprised of an outcome and features (predictors or covariates). A common scientific task is to separate important features (signals) from unrelated features (statistical noise) to facilitate modeling, learning, clinical diagnosis, and decision-making. Statistical models are selected for a variety of reasons: predictive ability, interpretability, ability to perform parameter inference, and ease of computation. A model’s set of features is called its “support” and the task of recovering the model’s true support from observed data is called “support recovery”. A desirable variable selection method will tend to return the set of true predictors - i.e. those features with truly non-zero coefficients - with high probability. Support recovery aids inference, because knowing the model’s true support benefits parameter estimation by reducing bias and improving efficiency. While an incorrectly specified model can sometimes have better predictive performance than a correctly specified model (\cite{shmueli2010explain}), having the correct support is essential for achieving optimal statistical inference (\cite{zhang2009some, shortreed2017outcome}).

Penalized likelihood procedures, originally optimized for prediction tasks, are widely used for variable selection. The lasso, an $\ell_1$ penalization method, produces models with strong predictive ability (\cite{tibshirani1996regression}). However, the lasso solution that maximizes predictive ability does not always lead to consistent support recovery (\cite{leng2006note, meinshausen2006high, shmueli2010explain, bogdan2015slope}). This is because noise variables are often included in the lasso solution that maximizes predictive ability (\cite{meinshausen2006high}). The adaptive lasso (AL), which introduces weights in the $\ell_1$ penalty, was proposed to resolve the issue that lasso solutions can be variable selection inconsistent (\cite{zou2006adaptive}). With clever choice of tuning parameters, and in large samples, the adaptive lasso can recover the true support with high probability and yield parameter estimates that converge properly (\cite{zou2006adaptive}). Smoothly clipped absolute deviation (SCAD) (\cite{fan2001variable}) and minimax concave penalty with penalized linear unbiased selection (MC+) (\cite{zhang2010nearly}) make use of distinctive piecewise linear thresholding functions to bridge the gap between the $\ell_0$ and $\ell_1$ algorithms. Both SCAD and MC+ seek to preserve large coefficients, like the $\ell_0$ penalty does, and shrink small coefficients, like the $\ell_1$ penalty does. While their variable selection properties have been well established, these methods are still not widely used in routine practice. 

All of the above approaches place a strong emphasis on predictive ability, at the cost of subsequent inference tasks. Because inference is an essential component of scientific investigations, a variable selection approach that balances prediction and inference tasks is highly desirable. Since traditional p-values do not reflect whether a variable is scientifically relevant or not (\cite{heinze2018variable}), we investigated whether using second-generation p-values (SGPV) (\cite{blume2018second, blume2019introduction}) would lead to good support recovery and subsequent parameter estimation and prediction. SGPVs emphasize scientific relevance in addition to statistical significance, and thus they are a good tool for screening out noise features and identifying the true signals in a set of candidate variables.

Following this idea, we propose a variable selection algorithm based on an $\ell_0$-Penalized regression with SGPVs (ProSGPV). The ProSGPV algorithm has a high support recovery rate and low parameter estimation bias, while maintaining good prediction performance even in the high-dimensional setting where $p>n$. In a series of comprehensive simulations and a real-world application, the ProSGPV algorithm is shown to be a viable alternative to, and often a noticeable improvement on, current variable selection standards such as AL, SCAD and MC+. While only linear models are discussed in this paper, forthcoming work will show that the ProSGPV approach generalizes to models of other classes, including logistic regression, Poisson regression, Cox proportional hazards model, etc.  
 
The structure of this paper is as follows. Section \ref{sec:backmat} provides a brief background. Section \ref{sec:2s} describes the proposed ProSGPV algorithm. Section \ref{sec:simulation} presents simulation studies comparing ProSGPV to AL, SCAD, and MC+ under various feature correlation structures and signal-to-noise ratios. Section \ref{sec:realw} illustrates the ProSGPV algorithm using a real-world data application. Section \ref{sec:concl} discusses the practical implications of the simulation results and some limitations of ProSGPV, and summarizes key findings in the paper.

\section{Background material}
\label{sec:backmat}

We review some fundamental ideas related to shrinkage, thresholding, inference, and prediction in the variable selection context to  facilitate subsequent discussions about ProSGPV. Readers familiar with standard variable selection notation, lasso (section \ref{subsec:lasso}), adaptive lasso (section \ref{subsec:al}), SCAD and MC+ (section \ref{subsec:scadmcp}), and second-generation p-values (section \ref{subsec:sgpv}) may skip to section \ref{sec:2s} for the development of the ProSGPV algorithm.

\subsection{Lasso}\label{subsec:lasso}
The lasso is an $\ell_1$ penalization procedure and one of the most widely used regularization methods for prediction modeling (\cite{tibshirani1996regression}). It reduces the feature space and identifies a subset of features that maximize predictive accuracy subject to a sparsity condition induced by the $\ell_1$ penalty. The set of features selected by lasso is called the active set. 

Let $\boldsymbol Y=(Y_1,Y_2,...Y_n)$ denote the response vector, $\boldsymbol X$ denote the $n\times p$ design matrix, and $\boldsymbol\beta\in\mathbb R^p$ denote the coefficient vector. $\lambda>0$ is a regularization parameter. $||\cdot||^2_2$ is the squared $\ell_2$-norm and $||\cdot||_1$ is the $\ell_1$-norm. Formally, the lasso solution is written as
\begin{equation} \label{eq:3}
\hat{\boldsymbol{\beta}}=\argmin_{\boldsymbol\beta} \{\frac{1}{2}||\boldsymbol Y-\boldsymbol{X\beta}||^2_2+\lambda||\boldsymbol  \beta||_1 \}
\end{equation}
The lasso is often used for variable selection because its solution encourages sparsity in the active set. However, even in the classical setting of a fixed $p$ and a growing $n$, the lasso active set tends to be different from the set of true signals. An exception to this is when true feature columns are roughly orthogonal to noise feature columns (\cite{knight2000asymptotics}), which unfortunately, is seldom seen in practice. \cite{wainwright2009sharp} improved the ability of the lasso solution to recover the true support under random Gaussian designs and showed that lasso can recover the true support when the effect size is sufficiently large and when no noise variables are highly correlated with true features. However, these conditions are strong and hard to apply in practice. In addition, even when they are met, there is no explicit way to implement the procedure because the shrinkage factor $\lambda$ that yields the correct support recovery is unknown (\cite{wang2013calibrating}). Lastly, the soft thresholding function in lasso shrinks large effects and results in biased parameter estimates that are ideal for prediction tasks, but not necessarily optimal for inference tasks.  

\subsection{Adaptive lasso}\label{subsec:al}

The adaptive lasso (AL) uses weights in the $\ell_1$ penalty to address the inconsistent variable selection property of the lasso (\cite{zou2006adaptive}). With the right shrinkage parameter, initial weights, and weight moments, the adaptive lasso can recover the true support with high probability while preserving prediction performance. Formally, the solution to the adaptive lasso is:  
\begin{equation} \label{eq:5}
\hat{\boldsymbol{\beta}}^{n}=\argmin_{\boldsymbol\beta} \{\frac{1}{2}||\boldsymbol Y-\boldsymbol{X\beta}||^2_2+\lambda_n||\boldsymbol{\hat \omega \beta}||_1 \}
\end{equation}
where $\boldsymbol{\hat\omega}=1/|\hat{\boldsymbol \beta}^*|^\gamma$. Here $\gamma>0$ is a tuning parameter and $\hat{\boldsymbol\beta}^*$ is any root-$n$-consistent estimator of the parameter $\boldsymbol\beta$, for example, an OLS estimator, or a lasso estimator.

\cite{zou2006adaptive} showed that AL has large-sample oracle (optimal) properties for support recovery and parameter estimation as $\lambda_n/\sqrt{n}\rightarrow 0$ and $\lambda_n n^{(\gamma-1)/2}\rightarrow\infty$. However, when the sample size is finite, it can be hard to find a combination of $\hat{\boldsymbol\beta}^*$, $\gamma$, and $\lambda_n$ such that the resulting active set matches the true support and the estimated coefficients have low bias.

\subsection{SCAD and MC+}\label{subsec:scadmcp}

SCAD and MC+ were designed to bridge $\ell_0$ and $\ell_1$ penalization schemes. As a result, both algorithms use nonconvex penalties. There are considerable advantages that come with using nonconvex penalization, such as a sparse solution and reduced parameter estimation bias, see \cite{fan2011nonconcave, fan2013asymptotic, zheng2014high, loh2015regularized}.

The penalty function in the SCAD corresponds to a quadratic spline function with knots at $\lambda$ and $\gamma\lambda$ (\cite{fan2001variable}). With proper choice of regularization parameters, SCAD can yield consistent variable selection in large samples (\cite{fan2001variable}). MC+ has two components: a minimax concave penalty (MCP) and a penalized linear unbiased selection (PLUS) algorithm (\cite{zhang2010nearly}). MC+ returns a continuous piecewise linear path for each coefficient as the penalty increases from zero (least squares) to infinity (null model). When the penalty level is set to $\lambda=\sigma\sqrt{(2/n)\log (p)}$, the MC+ algorithm has a high probability of support recovery and does not need to assume the strong irrepresentable condition (\cite{wainwright2009sharp}) that is required by lasso for support recovery (\cite{zhang2010nearly}). For visualization, Figure \ref{fig:1} displays the thresholding functions of $\ell_0$ and $\ell_1$ penalties, SCAD, and MC+ when the feature columns are orthogonal.

\subsection{Second-generation p-values}\label{subsec:sgpv}

Second-generation p-values (SGPV), denoted as $p_\delta$, were proposed for use in high dimensional multiple testing contexts (\cite{blume2018second, blume2019introduction}). SGPVs attempt to resolve some of the deficiencies of traditional p-values by replacing the point null hypothesis with a pre-specified interval null $H_0=[-\delta,\delta]$. The idea is to use the interval as a buffer region between “null” and “non-null” effects. The interval represents the set of effects that are scientifically indistinguishable or immeasurable from the point null due to limited precision or practicality. SGPV are essentially the fraction of data-supported hypotheses that are null, or nearly null, hypotheses.  

Formally, let $\theta$ be a parameter of interest, and let $I=[\theta_l,\theta_u]$ be an interval estimate of $\theta$ whose length is given by $|I|=\theta_u-\theta_l$. In this paper we will use a 95\% CI for $I$, but any type of the uncertainty interval can be used. If we denote the length of the interval null by $|H_0|$, then the SGPV $p_\delta$ is defined as 
\begin{equation}\label{eq:8}
p_\delta=\frac{|I\cap H_0|}{|I|}\times \max\left\{\frac{|I|}{2|H_0|},1\right\}  \end{equation}
where $I\cap H_0$ is the intersection of two intervals. The correction term $\max\{|I|/(2|H_0|),1\}$ applies when the interval estimate is very wide, i.e., when $|I|>2|H_0|$. In that case, the data are often inconclusive and the correction term shrinks the SGPV back to 1/2. As such, SGPVs indicate when data are compatible with null hypotheses ($p_\delta = 1$), or with alternative hypotheses ($p_\delta = 0$), or when data are inconclusive ($0 < p_\delta < 1$). 

By design, SGPVs emphasize effects that are scientifically meaningful as defined by exceeding a pre-specified effect size $\delta$. Empirical studies have shown that SGPVs have the potential for identifying feature importance in high dimensional settings (\cite{blume2018second, blume2019introduction}). This idea dovetails well with the natural tendency in variable selection to keep variables whose effects are above some threshold, say $\delta$. One extension here is that we will let the null bound $\delta$ shrink to zero at a pre-specified rate. This slight modification of the basic SGPV idea makes variable selection by SPGVs much more effective. Sensitivity to the choice of the null bound is assessed in Section \ref{subsec:nullbound}. 

\section{The ProSGPV algorithm}\label{sec:2s}

The ProSGPV algorithm is a two-stage algorithm. In the first stage, a candidate set of variables is acquired. In the second stage, an SGPV-based thresholding is applied to select variables from the candidate set that are meaningfully associated with the outcome.

\subsection{Steps}\label{subsec:2sstep}
The steps of the ProSGPV algorithm are shown below in Algorithm \ref{algo::2}.

\begin{algorithm}[H]
\caption{\label{algo::2} ProSGPV}
\begin{algorithmic}[1]
    \Procedure{ProSGPV}{$\boldsymbol X$, $\boldsymbol Y$}
    \State \textbf{Stage one}: Find a candidate set 
    \Indent
    \State Standardize all inputs (the outcome and features)
    \State Fit a lasso and find $\lambda_{\text{gic}}$ using generalized information criterion 
    \State Fit an OLS model on the lasso active set 
    \EndIndent
    \State \textbf{Stage two}: SGPV screening 
    \Indent
    \State Extract the confidence intervals of all variables from the previous OLS model   
    \State Calculate the mean coefficient standard error $\overline{SE}$ of standardized features
    \State Get the SGPV for each variable $k$ with $I_k=\hat\beta_k \pm 1.96\times SE_k$ and $H_0=[-\overline{SE}, \overline{SE} ]$ 
    \State Keep variables with SGPV of zero 
    \State Re-run the OLS model with selected variables on the original scale  
    \EndIndent
    \EndProcedure
\end{algorithmic}
\end{algorithm}

Note that the outcome and features are standardized except for the final step. Generalized information criterion (GIC) (\cite{fan2013tuning}) is used to find the shrinkage parameter $\lambda_\text{gic}$ that leads to a fully-relaxed lasso (\cite{meinshausen2007relaxed}) in the first stage. $\lambda$ could also be found through cross-validation, as there is evidence that a range of $\lambda$s will lead to the true support (\cite{fan2001variable, zou2006adaptive, wang2013calibrating, sun2019hard}). That adds to the flexibility of the algortihm. In the second stage, SGPVs are used to screen variables in the candidate set, where the null bound $\delta$ is derived from coefficient standard errors. Sensitivity to the choice of the null bound $\delta$ is assessed in Section \ref{subsec:nullbound}. We have implemented the ProSGPV algorithm in the \textbf{ProSGPV} \textsf{R} package, which is available from the Comprehensive R Archive Network (CRAN) at \url{https://CRAN.R-project.org/package=ProSGPV}.  

\subsection{Solution}\label{subsec:2ssol}

The solution to the ProSGPV algorithm $\hat{\boldsymbol\beta}^{\text{pro}}$ is 
\begin{equation}\label{eq:10}
\begin{gathered}
\hat{\boldsymbol\beta}^{\text{pro}}=\hat{\boldsymbol\beta}^{\text{ols}}_{|S}\in\mathbb R^p, \text{ where } \\
S=\{k\in C: |\hat\beta_k^{\text{ols}}|>\lambda_k \}, C=\{j\in\{1,2,...,p\}: |\hat{\beta}^{\text{lasso}}_j|>0\}   
\end{gathered}
\end{equation}
where $\hat{\boldsymbol\beta}^{\text{ols}}_{|S}$ is a vector of length $p$ with non-zero elements being the OLS coefficient estimates from the model with variables only in the set $S$, the final selection set. $C$ is the candidate set from the first-stage screening. $\hat\beta_j^{\text{lasso}}$ is the $j$th lasso solution evaluated at $\lambda_{\text{gic}}$ in the first stage. In the second stage, the cutoff is $\lambda_k=1.96\times SE_k+\overline{SE}$ and $\lambda_k$ is constant over $k$ when the features are all centered and standardized. In that case, the coefficient standard errors are identical.  

The ProSGPV algorithm is effectively a hard thresholding function. In the first stage, variables not selected by lasso are shrunk to zero. In the second stage, ProSGPV relaxes the coefficients and shrinks effects smaller than $\overline{SE}$ to zero while preserving large effects. Because this is a two-stage algorithm, there does not appear to be a simple closed-form solution for the implied thresholding without conditioning on the first stage. However, this would-be threshold, call it $\lambda_\text{new}$, tends to be larger than $\lambda_\text{gic}$ from lasso. The only routine exception to this is when data have weak signals or high correlation. But in that case, no algorithm can fully recover the true support (\cite{zhao2006model, wainwright2009information}) .

\begin{figure}[!ht]
\centering
\includegraphics[width=\textwidth]{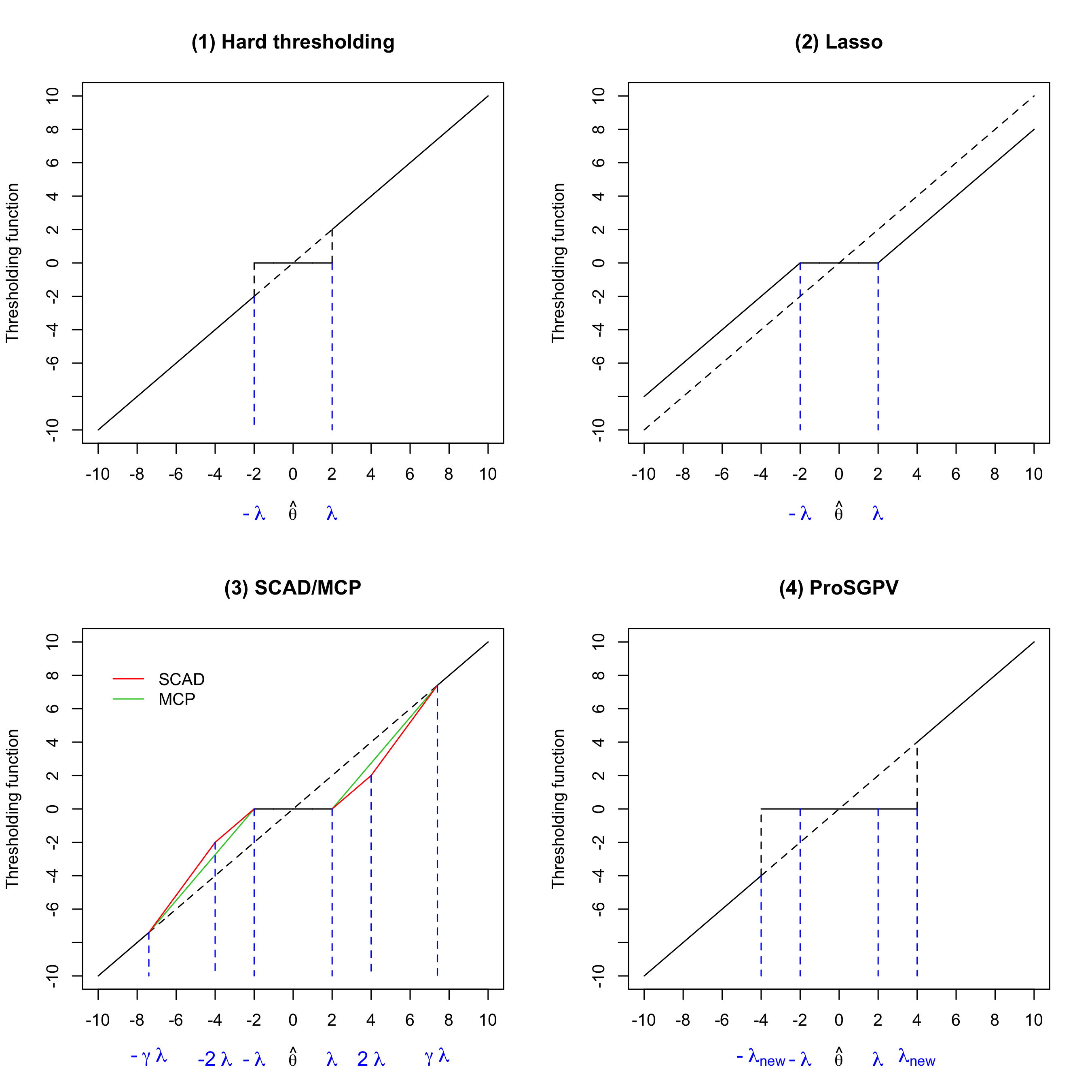}
\caption{\label{fig:1} Thresholding functions from five algorithms when features are orthogonal. }
\end{figure}

A visualization of thresholding functions for several penalization methods (assuming orthogonal features) is displayed in Figure \ref{fig:1}. The hard thresholding function in the panel (1) shrinks the coefficient estimates to zero when the effects are less than $\lambda$ and preserves them otherwise. The lasso in (2) shrinks small effects to zero and shrinks large effects by $\lambda$. SCAD and MCP in (3) bridge the gap between a hard thresholding function seen in (1) and a soft thresholding function in (2). When the coefficient is small ($|\hat\theta|\leq \lambda$), both methods have the same behavior as the lasso because the coefficient is shrunk to zero in all cases. When the coefficient is large ($|\hat\theta|\geq\gamma\lambda$), SCAD and MCP have the same behavior as the hard thresholding (no shrinkage is applied). What distinguishes SCAD and MCP is the shape of its thresholding function between $\lambda$ and $\gamma\lambda$. As mentioned earlier, ProSGPV amounts to a hard thresholding function whose cutoff is usually larger than $\lambda$.  

\subsection{Null bound}\label{subsec:nullbound}

The null bound in ProSGPV is set to be the average coefficient standard error, say $\overline{SE}$, from the OLS model on the lasso candidate set. Because of the scaling, this is equivalent to hard-thresholding variables whose absolute coefficients are below $1.96\times SE_k+\overline{SE}\approx 3\times\overline{SE}$. This is in line with variable selection ideas from literature. \cite{fan2006statistical} argued that in order to achieve optimal properties of variable selection, the amount of lasso shrinkage must be proportional to the standard error of the maximum likelihood estimates of coefficients. Intuitively, the interval null acts as a buffer zone to screen out effects that are likely false discoveries. By definition, the sampling distribution of false discoveries will be near the point null (since they are “false” discoveries) and the variance of this distribution shrinks at a rate proportional to the information in the sample. Hence, using the SE to delineate the smallest effect size of interest is natural. It is possible that a constant multiplier of the SE might yield a better Type I-Type II error tradeoff, but after trying some obvious variations we did not find anything better.

A sensitivity analysis on the choice of the null bound was conducted and is summarized in Supplementary Figure 1. We compared the support recovery performance of ProSGPV using different null bounds when signal-to-noise ratio (SNR) is medium or high and when $n>p$. Choices of null bounds include the original bound $\overline{SE}$, $\overline{SE}\times \sqrt{\log (n/p)}$, $\overline{SE}/ \sqrt{\log (n/p)}$, $\hat\sigma/12$, and 0. When the null bound is constant, e.g., $\hat\sigma/12$, the support recovery performance is poor. When the null bound is scaled by $\sqrt{\log (n/p)}$, performance appears to be slightly improved in the high correlation case, but, importantly, is inferior in all other cases. When the null bound is set at 0, ProSGPV amounts to selecting variables using traditional p-values. In this case, the support recovery performance is expectedly poor even when SNR is high, because the null bound of 0 leads to many false positives (\cite{kaufman2014does,janson2015effective}). When $p>n$, the above observations hold because the null bound is calculated from a model with a reduced number of features (same order as $s<<p$, where $s$ is the number of true signals). This sparsity assumption is necessary for successful high-dimensional support recovery (\cite{meinshausen2006high,zhao2006model,wainwright2009information}). Hence, allowing the null bound, which acts as a thresholding function, to shrink at a $\sqrt n$-rate, appears to offer the best performance across the widest range of scenarios.

\subsection{Example}\label{subsec:2sexam}

\begin{figure}[!ht]
\centering
\includegraphics[width=0.95\columnwidth]{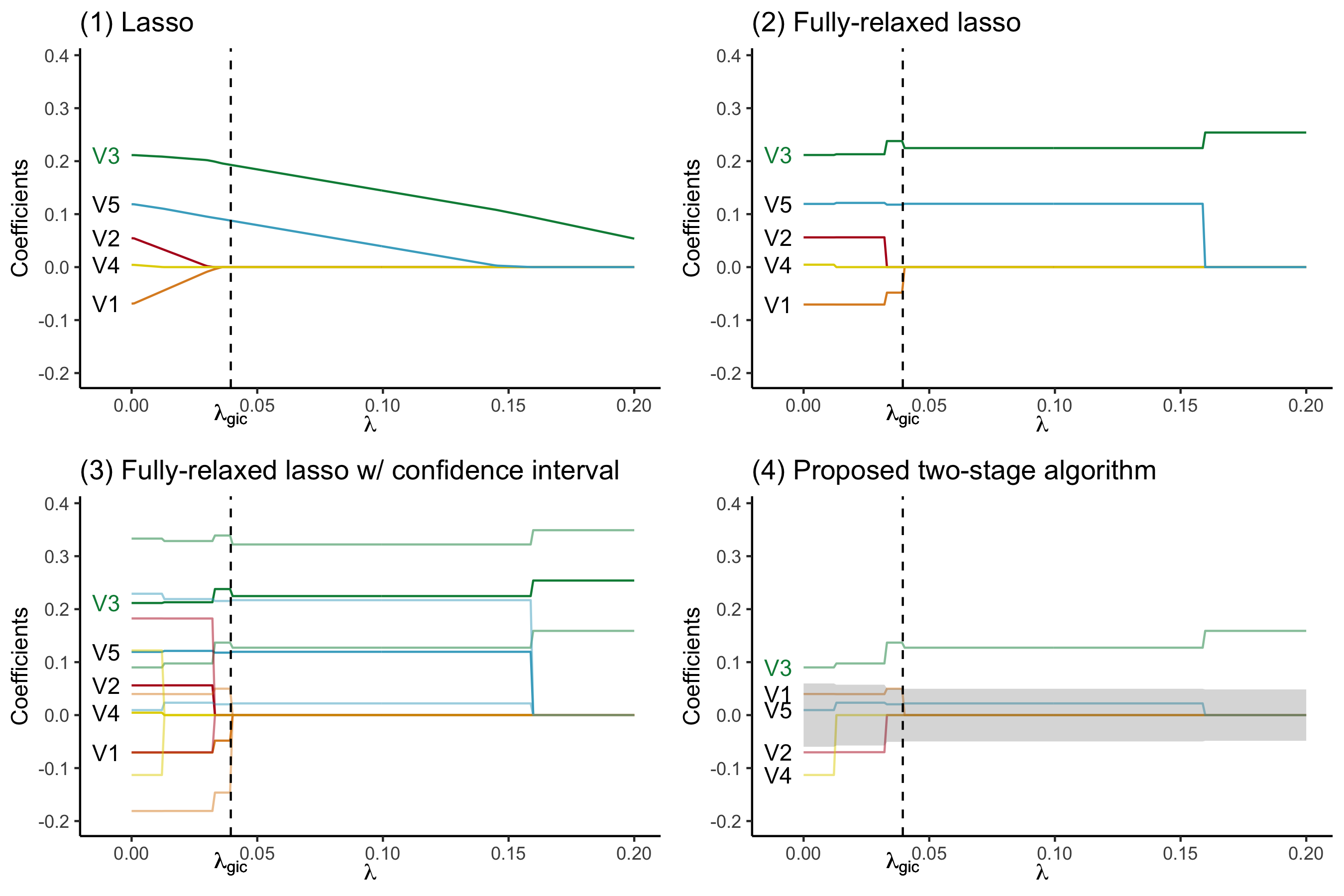}
\caption{\label{fig:2} Illustration of the ProSGPV algorithm. Panel (1) presents the colored lasso solution path where the vertical dotted line is the $\lambda_{\text{gic}}$. Panel (2) shows the fully-relaxed lasso path with point estimates only. Panel (3) shows the same path plus 95\% confidence intervals in light colors. Panel (4) is the proposed two-stage algorithm’s selection path. The shaded area is the null region and only the 95\% confidence bound that is closer to zero is shown for each variable.}
\end{figure}

Figure \ref{fig:2} shows the effect of the ProSGPV algorithm on the regression coefficients in our simulated setting. Suppose that the true data-generating model is $\boldsymbol y=\boldsymbol{X\beta}+\boldsymbol\epsilon$ where $\boldsymbol y$ is a vector of length 400. The design matrix $\boldsymbol X$ has five columns with mean zero and covariance matrix $\Sigma_{i,j}=0.5^{|i-j|}$. The coefficient vector $\boldsymbol\beta$ is zero everywhere except $\beta_3=0.28$. The errors are i.i.d. $N(0,1)$. We see in Figure \ref{fig:2} that the ProSGPV algorithm succeeds by selecting V3 whereas the lasso and relaxed lasso select V3 and V5 at $\lambda_{\text{gic}}$. 

\subsection{Similar algorithms from the literature}\label{subsec:2ssim}

Other two-stage algorithms have been proposed for pre-screening features (\cite{meinshausen2009lasso,zhang2009some,wasserman2009high,zhou2009thresholding,zhou2010thresholded,sun2019hard,weng2019regularization,wang2020bridge}). \cite{meinshausen2009lasso} proposed a two-stage thresholded lasso, where a lasso model is fit and features are kept if they pass a data-dependent coefficient threshold. Because of this, the resulting coefficient estimates are biased even when the correct support is recovered. \cite{wasserman2009high} proposed using variable selection methods (lasso, marginal regression, forward stepwise regression, etc.) with cross-validation to pre-screen candidate variables before using Bonferroni corrected t-tests to identify and remove noise features. Wasserman’s method controls the Type I error rate across all features, but pays a higher price in false negatives. ProSGPV, however, allows the Type I error rate to shrink towards zero and yields fewer false positives (See Supplementary Figure 3). \cite{zhang2009some} identified relevant and irrelevant features from lasso in the first stage and fit another $\ell_1$-penalized regression using only irrelevant features afterwards. However, \cite{zhang2009some} emphasizes parameter estimation and neglects support recovery. In addition, their algorithm needs to run multiple cross-validations while ProSGPV uses GIC to tune $\lambda$ and is therefore much faster to compute. \cite{sun2019hard} proposed the hard thresholding regression (HRS). When lasso is used to derive initial weights, the HRS reduces to the fully relaxed lasso, which is the first stage of our two-stage ProSGPV algorithm. Unlike our algorithm, HRS keeps all variables that survive the first stage. Lastly, \cite{zhou2009thresholding, zhou2010thresholded} used lasso or the Dantzig selector to pre-screen and then used a fully relaxed model on thresholded coefficients with a data-driven bound; \cite{weng2019regularization} selected important variables and penalized only the unselected variables for the final variable selection; \cite{wang2020bridge} used a bridge regression in the first stage and thresholded variables in the second stage. 

\subsection{Special case: one-stage ProSGPV algorithm}\label{subsec:specialcase}

When $\lambda_{\text{gic}}$ is replaced with zero in the first stage of lasso, ProSGPV reduces to a one-stage algorithm. That amounts to calculating the SGPV for each variable in the full OLS model and selecting ones that are above the threshold. The one-stage ProSGPV is faster to compute, as no lasso solution path is required. However, it does not appear to be variable selection consistent in the limit, and its inferential performance is inferior to that of the two-stage ProSGPV when data do not contain strong signals or features are highly correlated. Moreover, it is not applicable when $p>n$, i.e., when the OLS model is not identifiable. For completeness, the support recovery performance of the one-stage algorithm can be found in Supplementary Figure 2. Its performance is very close to the two-stage algorithm when explanatory variables are independent.

\subsection{Summary}\label{subsec:2ssumm}

The ideas behind the ProSGPV algorithm are intuitive: exclude small effects using a data-dependent threshold for noise and keep large effects. ProSGPV is essentially an $\ell_0$-penalized regression. Unlike the $\ell_1$ penalty, $\ell_0$ optimization is nonconvex, so it is harder to compute and less popular in practice. However, our algorithm avoids enumerating all possible combinations of variables by leveraging the lasso solution in the first stage and threshold effects with an explicit bound afterwards. That translates into less computational cost than other convex optimization algorithms (as seen in Supplementary Figure 6). ProSGPV can also be thought of as a variation of the thresholded lasso with refitting. \cite{van2011adaptive} showed that the thresholded lasso with refitting requires less severe minimal signal conditions for successful support recovery than adaptive lasso. While lasso is used in the first stage screening, other variable selection methods, such as Sure Independence Screening (SIS) (\cite{fan2008sure}), can be used there. This adds the /flexibility to our algorithm. Lastly, in terms of post-selection inference, the point estimates and corresponding confidence intervals derived from our algorithm are best when the selected model matches the true underlying model. Even when ProSGPV misses true signals, those missed variables often have small effects, which results in minimal impact on the inference of the other larger effects.

\section{Simulation studies}\label{sec:simulation}

Extensive simulation studies were conducted to evaluate the inferential and prediction performance of the ProSGPV algorithm and compare it to existing methods. We investigated both traditional $n>p$ and high-dimensional $p>n$ settings.   

\subsection{Design}\label{subsec:design}

The simulation setup is motivated by similar investigations such as \cite{hastie2020best}. We set sample size $n$, dimension of explanatory variables $p$, sparsity level $s$ (number of true signals), true coefficient vector $\boldsymbol\beta_0\in\mathbb R^p$, autocorrelation level $\rho$ within explanatory variables, and signal-to-noise ratio (SNR) $\nu$. 

In the traditional $n>p$ setting, $p$ is fixed at 50 and $n$ ranges from 100 to 2000 with an increment of 50. The number of true signals $s$ is fixed at 10. In the high-dimensional setting, $n$ is fixed at 200 and $p$ ranges from 200 to 2000 with an increment of 20. Here, the number of true signals is fixed at 4. $\boldsymbol\beta_0$ has $s$ non-zero values equally-spaced between one and five, at random positions, and the rest are zero. The coefficients are half positive and half negative. $\rho$ can take the value of 0 (independent), 0.35 (medium autocorrelation), and 0.7 (high autocorrelation). SNR is defined as $\text{SNR}= Var(f(x))/Var(\epsilon)$, where data are generated from a probabilistic distribution. SNR take the value of 0.7 (moderate SNR), and 2 (high SNR) (\cite{hastie2020best}).    

We evaluated the performance of each algorithm using standard metrics: support recovery rate, Type I error rate, power, false discovery rate, false non-discovery rate, along with the mean absolute error (defined below) for parameter estimation, prediction accuracy in a separate test set, and running time. See Supplement Table 1 for detailed definitions of the metrics for inference.  

\textbf{Step 1}: Draw $n$ rows of the matrix $\boldsymbol X\in\mathbb R^{n\times p}$ i.i.d. from $N_p(0,\boldsymbol\Sigma)$, where $\boldsymbol\Sigma\in \mathbb R^{p\times p}$ has entry $(i,j)$ equal to $\rho^{|i-j|}$.

\textbf{Step 2}: Generate the response vector $\boldsymbol Y\in\mathbb R^{n}$ from $N_{n}(\boldsymbol {X\beta}_0,\sigma^2 \boldsymbol I)$, with $\sigma^2$ defined to meet the desired SNR level $\nu$, i.e., $\sigma^2=\boldsymbol\beta_0^T \boldsymbol\Sigma\boldsymbol\beta_0/\nu$. 

\textbf{Step 3}: Run SCAD, MC+, AL, and ProSGPV on the training set with $n$ observations; record the active set from each algorithm; compute evaluation metrics in Supplementary Table 1 plus capture rate of the exact true model, absolute bias in parameter estimation, and running time; use a separate test set to compute prediction accuracy. Note that the test set was generated in Step 1, and set aside for later use by inflating the target sample size $n$.

\textbf{Step 4}: Repeat the previous steps 1000 times and aggregate the results. 

SCAD was implemented using the \textbf{ncvreg} package in \textsf{R} and $\gamma$ was fixed at 3.7, MC+ was implemented using the \textbf{plus} package. Adaptive lasso was implemented using the \textbf{glmnet} package and the initial weights are the inverse of absolute value of lasso estimates. For a fair comparison, GIC was used to select $\lambda$ in all algorithms. The ProSGPV algorithm was implemented using the \textbf{ProSGPV} package. The \textsf{R} code to replicate simulation results can be found at \url{https://github.com/zuoyi93/r-code-prosgpv-linear}.

\subsection{Results and findings}\label{subsec:simresults}

\begin{figure}[!ht]
\centering
\includegraphics[width=0.95\columnwidth]{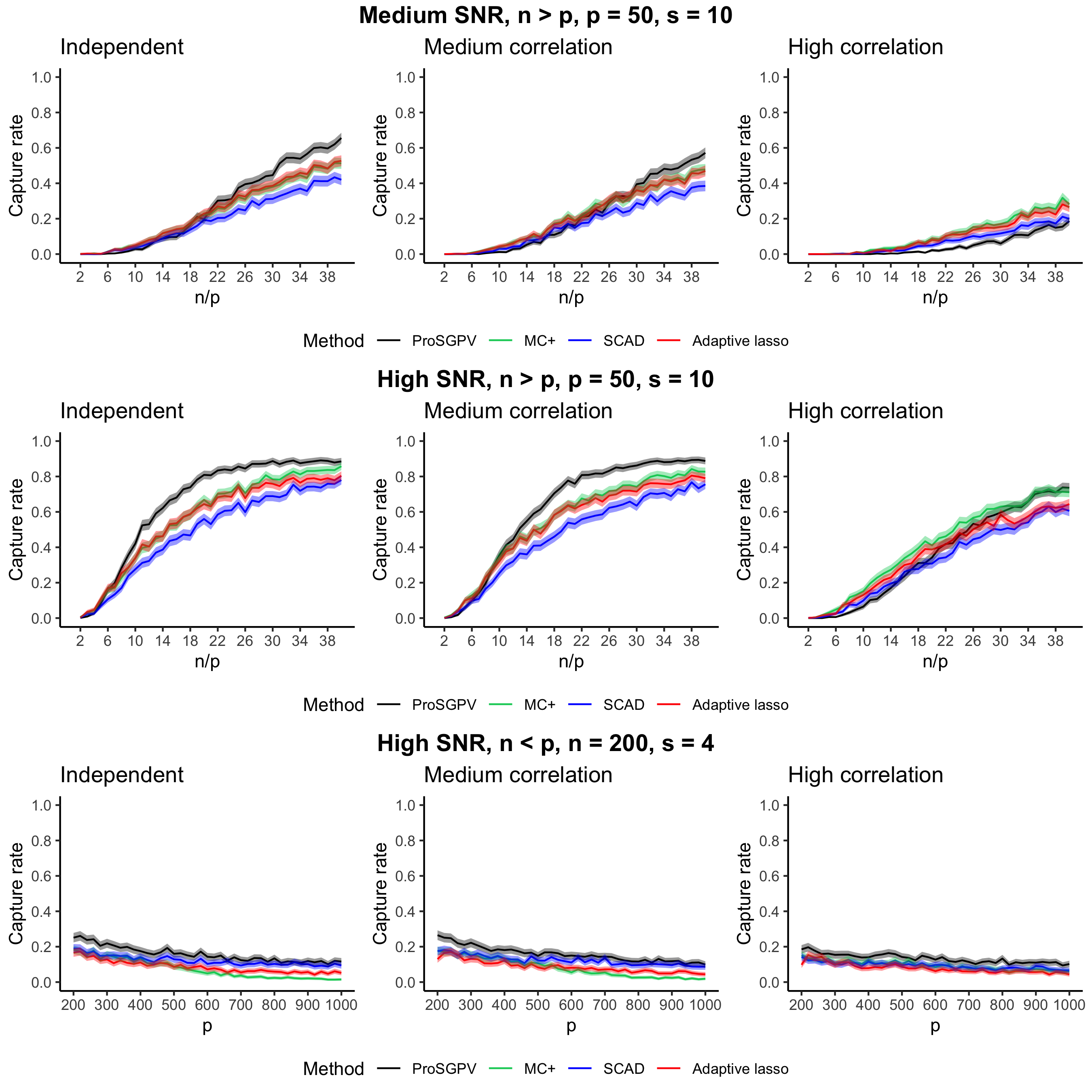}
\caption{\label{fig:3} Capture rate of the exact true model under combinations of autocorrelation level, signal-noise-ratios, and $(n,p,s)$. In each panel, one algorithm has a colored solid line representing the average capture rate surrounded by the shaded 95\% Wald interval over 1000 simulations.}
\end{figure}

\begin{figure}[!ht]
\centering
\includegraphics[width=0.95\columnwidth]{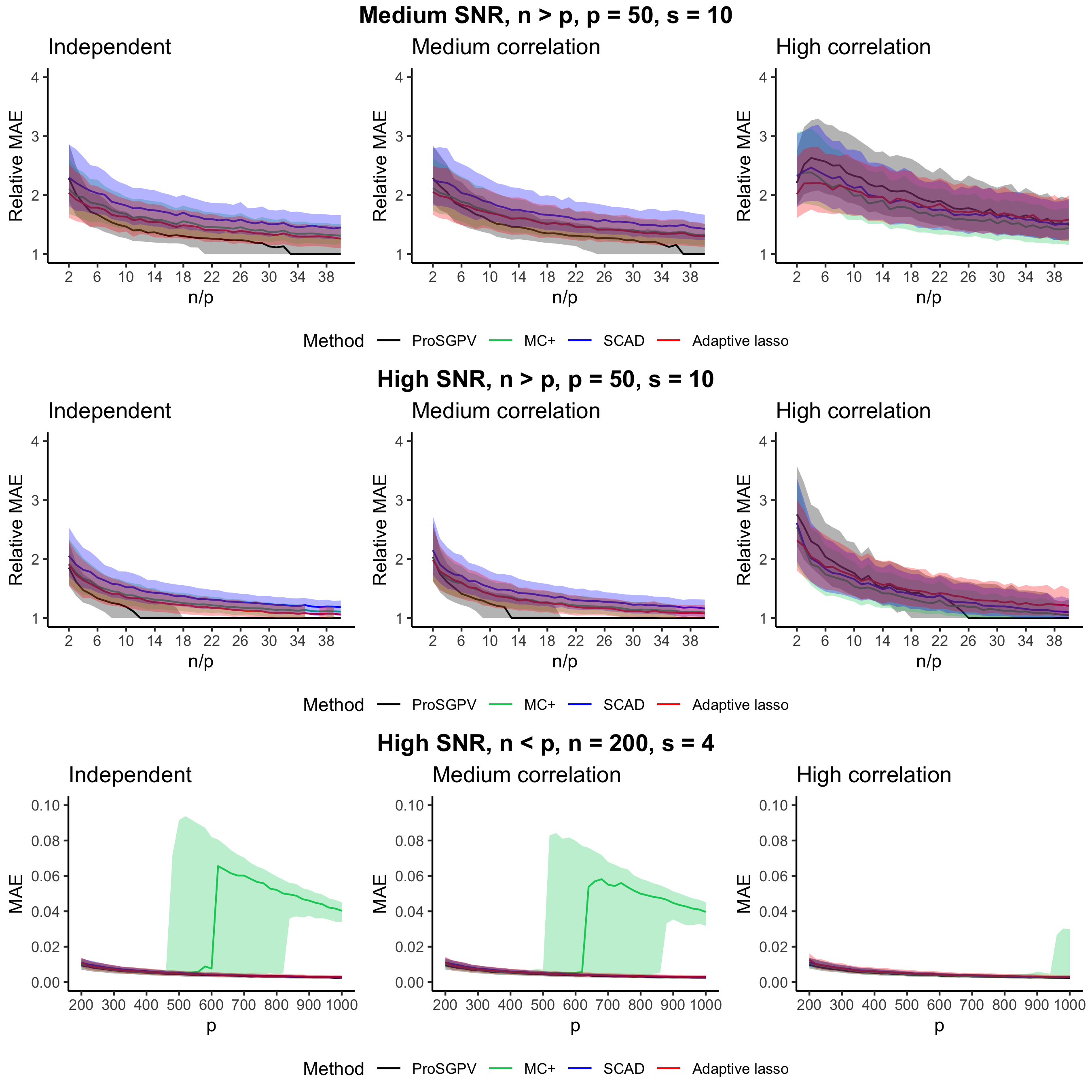}
\caption{\label{fig:4} Parameter estimation error of all algorithms under combinations of autocorrelation level, signal-to-noise ratio, and $(n, p, s)$. In each panel, one algorithm has a colored solid line representing the median (relative) mean absolute errors surrounded by the shaded first and third quartiles over 1000 simulations.  }
\end{figure}

We recorded whether or not each algorithm captured the exact true model in each iteration and compared the average capture rates over 1000 iterations in Figure \ref{fig:3}. We also compared the mean absolute error (MAE) of all coefficient estimates, defined as $\frac{1}{p}\sum_{j=1}^p\vert\hat\beta_j-\beta_{0,j} \vert$, in Figure \ref{fig:4}, where $\beta_{0,j}$ is the $j$th true coefficient. We compared the prediction accuracy of each algorithm, as measured by root mean square error (RMSE) in an independent test set in Figure \ref{fig:5}. Power and Type I error rates are presented in Supplementary Figure 3. False discovery proportions (pFDR) and false non-discovery proportions (pFNR) are presented in Supplementary Figure 4. The effect of different parameter tuning methods on MC+ is illustrated in Supplementary Figure 5. The comparison of computation time is shown in Supplementary Figure 6. 

In Figure \ref{fig:3}, capture rates of the exact true model are compared under combinations of SNR and autocorrelation levels within the design matrix, when both $n>p$ and $n<p$. 

When $n>p$, ProSGPV’s support recovery rate increases as $n$ grows. It generally has the highest support recovery rate except when the SNR is medium and correlation is high. MC+ and AL have similar capture rates, while SCAD is the worst among the four. When $p>n$, support recovery rates are low for all methods and decrease as $p$ increases in the data. ProSGPV again is the highest, followed by SCAD, AL, and MC+. We investigated factors driving the support recovery performance in Supplementary Figure 3 and 4. When $n>p$, we see that all algorithms have decreasing Type I error rates, pFDR, pFNR, and increasing power. When $p>n$ and data are not highly correlated, GIC-based MC+ has notably higher pFDR than the others, indicating that it overfits the training data and includes many noise variables.

Mean absolute error (MAE) is used to assess the parameter estimation error. When $n>p$, we used relative MAE which is defined as the ratio of an algorithm’s MAE to that of the OLS model with only true features. A good estimator would have an asymptotic relative MAE of one. When $p>n$, absolute MAE is used because no OLS fit is possible. Figure \ref{fig:4} displays the median (relative) MAE of four algorithms under various scenarios. The shading shows the first and third quartiles of the empirical (relative) MAE distribution.

In both $n>p$ and $n<p$ cases, ProSGPV has the lowest parameter estimation error. This should not be surprising for sparse settings with well-defined signals, as ProSGPV is effectively an $\ell_0$ penalization derivative and $\ell_0$ penalization drops small effects while keeping large ones. \cite{johnson2015risk} showed that the parameter estimation risk of $\ell_0$-penalized regression can be infinitely better than that of the $\ell_1$-penalized regression under certain conditions and this is a practical example. The shape of the relative MAE from ProSGPV generally follows what would be expected from a rate of $\sqrt{\log (n) /n}$. This rate matches the ideal rate of parameter estimation in the optimal model from any hard thresholding function, as suggested by Theorem 1 of \cite{zheng2014high}. When $n>p$, AL and MC+ have very close performance and SCAD has the slowest rate of convergence. When $n<p$, the order stays the same for all except MC+. As $p$ passes 600, MC+ with GIC-based tunning selects more noise variables in the model and the parameter estimation performance is compromised. This can be remedied by using a universal $\lambda=\sigma\sqrt{(2/n)\log p}$ in MC+ (see Supplementary Figure 5). \cite{fan2013tuning} argued that MC+ has the same performance as SCAD when GIC is used. However, in their setting, $s$, $p$, and $n$ are allowed to grow together. In our case, $s$ is fixed at 4, $n$ is fixed at 200, and only $p$ grows.

In Figure \ref{fig:5}, the prediction RMSE is calculated in an independent test set (40\%) using models built with a training set (60\%). Again, when $n>p$ the relative RMSE is used while when $n<p$ the absolute RMSE is used. Relative RMSE is defined as the ratio of the prediction RMSE from one algorithm to that from the OLS model with true signals only.

\begin{figure}[!ht]
\centering
\includegraphics[width=0.9\columnwidth]{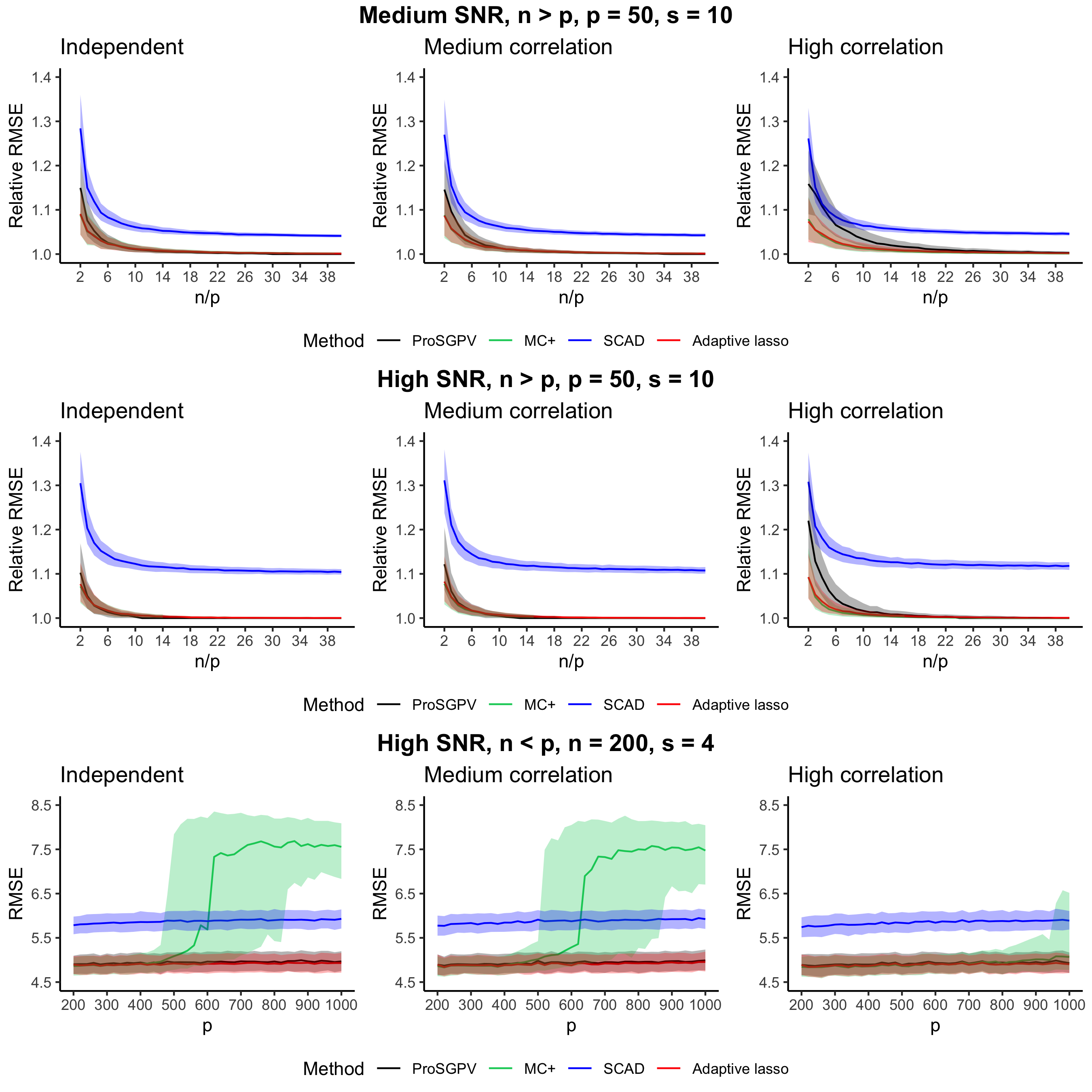}
\caption{\label{fig:5} Comparison of prediction accuracy of all algorithms under combinations of autocorrelation level, signal-to-noise ratio, and $(n,p,s)$. Median (relative) root mean square errors are surrounded by their first and third quartiles over 1000 simulations. }
\end{figure}

When $n>p$, all algorithms have worse prediction performance than the true OLS model unless $n$ is really large. But their prediction RMSEs converge to the true OLS RMSE from above as $n$ increases. While ProSGPV is not optimized for prediction tasks, its predictive ability quickly catches up with other algorithms when $n/p>6$. When $n<p$, ProSGPV and AL have the best performance followed by SCAD. SCAD can have better prediction performance when $\lambda$ is selected by cross-validation. However, in that case, its support recovery is worse than that from the GIC-based SCAD. MC+ has much higher prediction error than the others when $p>600$. That is because $\lambda$ selected by GIC leads to a dense model which includes many noise variables. The overfitted model has poor prediction performance in an external data set. However, this can be remedied by using a universal $\lambda$ in MC+, as shown in Supplementary Figure 5.

In Supplementary Figure 6, the running time in seconds from all algorithms are compared. The computing environment was 2.6 GHz Dual-Core Intel Core i7 processor and 32 GB memory. ProSGPV and AL have the shortest computation time, followed by SCAD. MC+ is more time-consuming when data are highly correlated, or when $n<p$.

\section{Real-world example}\label{sec:realw}

We illustrate our approach using the Tehran housing data (\cite{rafiei2016novel}), which was high SNR ($R^2= 0.98$) in the OLS model with all variables. We also explored the medium SNR case by removing potentially redundant variables until $R^2=0.4$. The Tehran housing data are available as a data object \textsf{t.housing} in the \textbf{ProSGPV} package. The data set contains 26 features and 372 records (see Supplementary Table 2 for the variable description). The goal is to predict the sale price (variable 9 or V9). The explanatory variables consist of seven project physical and financial variables, 19 economic variables, all at baseline. Clustering and correlation patterns are displayed in Supplementary Figure 7. We see that several explanatory variables form prominent clusters and that there is high pairwise correlation among the features. In particular, the price per square meter of the unit at the beginning of the project (V8) has high correlation ($\rho=0.98$) with the sale price (V9). 

We repeatedly split the data into a training set (70\%) and a test set (30\%). We applied AL, SCAD, MC+, and ProSGPV algorithms on the training set ($n$=260) with all the covariates. Prediction RMSE was calculated on the test set ($n$=112). We summarized the sparsity of the solutions (Supplementary Figure 8) and prediction accuracy (Supplementary Figure 9) over 1000 training-test split repetitions. SCAD overfits the training data and has the largest selection set. AL yields the sparsest model followed by ProSGPV. GIC-based MC+ yields a constant model size. Regarding the prediction performance, ProSGPV has the lowest median prediction error closely followed by AL and MC+, while SCAD has the largest test error because of overfitting. All algorithms select duration of construction (V7) and initial price per square meter (V8) with high frequency, and there is no consensus as for which other variables to include because of high correlation and clustering. 

To refine the analysis, we removed variables that had an absolute correlation with the outcome of 0.45 or greater. The remaining covariates explain 40\% of the variability in the response, which represents medium SNR. Of the remaining nine variables, MC+ always selects zero variables. ProSGPV selects four or five variables with high frequency. AL selects six or seven variables with high frequency. SCAD selects more variables than ProSGPV and AL. ProSGPV, SCAD and AL have similar prediction performance, while MC+ has worse performance, due to the null model it selects. The common selected variables include total floor area of the building (V2), lot area (V3), the price per square meter of the unit at the beginning of the project (V8), and the number of building permits issued (V11). The \textsf{R} code to replicate the results is available at \url{https://github.com/zuoyi93/r-code-prosgpv-linear}.

\section{Practical implications, limitations, and comments}\label{sec:concl}

A naïve way to perform variable selection is to screen variables by p-values. Such methods include forward selection, backward selection, and stepwise selection (\cite{efroymson1966stepwise}). However, these methods have serious drawbacks. They have poor capture rates of the true underlying model (\cite{wang2009forward, kozbur2018sharp}) and larger effective degrees of freedom (\cite{kaufman2014does, janson2015effective}). In addition, the standard errors of the coefficient estimates are too small, which leads to over-optimistic discoveries (\cite{harrell2015regression}). Better approaches do exist, but they are more complex, require specialized software, and are not fully adopted in routine applied practice. However, SGPVs offers a simple and effective option. It exhibits excellent statistical properties in both inference and prediction tasks without increased computation time.

Our simulation studies reinforce the notion that a model with good prediction ability does not necessarily lead to good inference. Comparing Figure \ref{fig:3} with Figure \ref{fig:5}, we see that models optimized for prediction tend not to be optimized for inferential tasks, even when we use a different parameter tuning approach for each algorithm. This corroborates findings in the literature (\cite{leng2006note, meinshausen2006high, wasserman2009high, zheng2014high, giacobino2017quantile, shortreed2017outcome}). This statement is important and bears repeating: models optimized for prediction tasks do not necessarily support good inference. Similar observations can be found by comparing the parameter estimation in Figure \ref{fig:4} with prediction performance in Figure \ref{fig:5}. ProSGPV does a better job by yielding a model that is primed for inference and also has good prediction properties.

There is a link between SNR and the proportion of variance explained (PVE).
\begin{equation}\label{eq:11}
\text{PVE}(f)=1-\frac{\mathbb E (y-f(x)^2)} {\text{Var}(y)}=1-\frac{\text{Var}(\epsilon)}{\text{Var}(y)}=\frac{\text{SNR}}{1+\text{SNR}}
\end{equation}
where $f$ is the mean function, and $x$ is independent from $\epsilon$. Practically, when the $R^2$ is around 0.40 in the full model, which is equivalent to a medium SNR in our simulation, ProSGPV has a comparable performance in support recovery and slightly better parameter estimation in large $n$; when the $R^2$ is above 0.66, which corresponds to a high SNR, ProSGPV has superior inference properties than the other algorithms.  

There are some limitations. Sensitivity to tuning parameter specification is an issue for both implementation and generalizability of results. However, we found the findings to be fairly robust to tuning parameter specification. In results not shown here, we repeated the experiment using each algorithm’s preferred method for choosing a tuning parameter and the general ordering of results remained stable. Another limitation is when the design matrix has high within-correlation, which is a challenging problem for any algorithm. Not unexpectedly, ProSGPV does not do well in support recovery and its parameter estimation and prediction performance suffers. We also note that the exact threshold function of the two-stage ProSGPV algorithm is difficult to conceptualize, as the null bound in the fully relaxed lasso has a different feature space than the full feature space. We are actively working on formulating solutions for the two-stage algorithm. 

Despite these relatively minor limitations, the ProSGPV algorithm looks very promising. It gives up little in terms of prediction, and offers improved support recovery and parameter estimation properties compared to the class of standard procedures currently in use today. Also, the ProSGPV algorithm does not depend on tuning parameters that are hard to specify. It is fair to say that unlike traditional p-values, second-generation p-values can be used for variable selection and subsequent statistical inference.

\bibliographystyle{plainnat}
\bibliography{refs}
\end{document}